\documentclass[preprint,nofootinbib]{revtex4}
\usepackage{amsthm, amscd, amsfonts,graphicx, amsmath,amssymb, subfig,environ,color,hyperref}
\usepackage{tensor}
\usepackage{filecontents}

\newcommand{\ba}{\begin{align}}
\newcommand{\ea}{\end{align}}
\newcommand{\bq}{\begin{equation}}
\newcommand{\eq}{\end{equation}}
\newcommand{\barr}{\begin{eqnarray}}
\newcommand{\earr}{\end{eqnarray}}

\newcommand{\di}{\, \mathrm{d}}

\newcommand{\rot}{\rho_{tot}}
\newcommand{\Pot}{P_{tot}}
\newcommand{\cc}{\cite}

\def\f{\frac}

\def\mR{\mathcal{R}}
\def\mH{\mathcal{H}}

\begin{document}

\title{Running of the spectral index in deformed matter bounce scenarios with
    Hubble-rate-dependent dark energy}

\author{M. Arab \footnote{Email:\text{moarab19@gmail.com}}~~and~~ A. Khodam-Mohammadi\footnote{Email:\text{khodam@basu.ac.ir (corresponding author)}}
}

\affiliation{Department of Physics, Faculty of Science, Bu-Ali Sina
University, Hamedan 65178, Iran}

\begin{abstract}
As a deformed matter bounce scenario with a dark energy component,
we propose a deformed one with running vacuum model (RVM) in which
the dark energy density $\rho_{\Lambda}$ is written as a power
series of $H^2$ and $\dot H$ with a constant equation of state
parameter, same as the cosmological constant, $w=-1$. Our results in
analytical and numerical point of views show that in some cases same
as $\Lambda$CDM bounce scenario, although the spectral index may
achieve a good consistency with observations, a positive value of
running of spectral index ($\alpha_s$) is obtained which is not
compatible with inflationary paradigm where it predicts a small
negative value for $\alpha_s$. However, by extending the power
series up to $H^4$, $\rho_{\Lambda}=n_0+n_2 H^2+n_4 H^4$, and
estimating a set of consistent parameters, we obtain the spectral
index $n_s$, a small negative value of running $\alpha_s$ and tensor
to scalar ratio $r$, which these reveal a degeneracy between
deformed matter bounce scenario with RVM-DE and inflationary
cosmology.
\end{abstract}
\maketitle



\section{introduction}
The idea of bouncing cosmology, mainly was suggested for replacing
the big bang singularity to a non-singular cosmology. More recent
observations of cosmic microwave background (CMB) give us some
evidence in which the scalar perturbations is nearly scale-invariant
at the early universe \cite{Hinshaw:2012aka,Ade:2013zuv}. Although
the inflationary scenario is the most currently paradigm of the
early universe and can solve several problems in standard big bang
cosmology, it faced with two basically problems. One key challenge
is the singularity problem before the beginning of inflation, which
is arisen from an extent of the Hawking-Penrose singularity theorems
which show that an inflationary universe is geodesically past
incomplete and it cannot reveal the history of the very early
universe \cite{Hawking:1969sw,Borde:1993xh}.

The second one is the \textit{trans-Planckian} problem which reveals
that the wavelength of all scales of cosmological interest today
originate in sub-Planckian values where the general relativity and
quantum field theory is broken down. Therefore, it leads to
important modifications of the predicted spectrum of cosmological
perturbations \cite{Martin:2000xs} (more details are referred to a
good informative review \cite{Brandenberger:2016vhg}). These
problems however have been avoided in the bouncing cosmology
\cite{Brandenberger:2016vhg}. At a bounce time ($t=0$), the space
gets a non-vanishing volume and also the wavelength of cosmological
perturbations is minimum in which their values correspond to the end
of inflation in cosmology. Due to this fact, the bouncing scenario
is usually considered as an alternative to inflationary
cosmology\cite{Brandenberger:2012zb}.

In light of cosmological perturbations, three familiar classes of
bouncing model which are differences in contracting phase have been
introduced. One of the most interested is a matter bounce scenario
\cite{Finelli:2001sr}. The others are Pre-Big-Bang
\cc{Gasperini:1992em} or Ekpyrotic \cc{Khoury:2001wf} type, matter
Ekpyrotic-bounce \cc{Cai:2014bea}, matter bounce inflation scenario
\cc{Cai:2014bea}, and string gas cosmology \cc{Brandenberger:1988aj,
Nayeri:2005ck}. In matter bounce scenario which have been widely
discussed in literatures
\cite{Lin:2010pf,WilsonEwing:2012pu,Cai:2015vzv,Cai:2016hea}, some
authors have considered one or two scalar fields \cite{Cai:2013kja},
others work with a semi matter (a matter with a dark energy
component) \cite{Cai:2014zga}, and many efforts have been done in
modified gravity and scalar tensor gravity
\cite{Bamba:2015uma,Odintsov:2014gea,Haro:2015zta,Boisseau:2016pfh}.
In all of them the dynamical behavior is described by loop quantum
cosmology (LQC)
\cite{Ashtekar:2011ni,Ashtekar:2007tv,Bojowald:2008ik,Cailleteau:2012fy,Quintin:2014oea,Cai:2011tc,Cai:2011zx,Amoros:2013nxa,Qiu:2013eoa,deHaro:2014kxa,Cai:2014zga}
around bouncing point which is arisen from quantum gravity in high
energy physics. Despite the success of LQC in non-singular bounce
cosmology, it is important to note that the dynamical mechanisms
that trigger non-singular bounce at high energy scales are not
always provided by the LQC. For instance in \cite{Cai:2012va} and
\cite{Cai:2013vm}, one can find an effective field theory model
including the Horndesky operators to give rise to a non-singular
bounce without much pathologies; additionally, the curvature
corrections appeared at high energy scales can also yield a
non-singular bounce, such as in \cite{Mukhanov:1991zn,
Brandenberger:1993ef} and very recently revisited in
\cite{Yoshida:2017swb}.

The LQC is also applied around turning point in a cyclic universe
scenario, where the universe goes to a contraction after an
expansion phase \cite{Sami:2006wj}. Although in many models of
matter bounce the power spectral index of cosmological perturbation
may be consistent with observations, they often obtain a positive
running of scalar spectral index ($\alpha_s$) which may be
irreconcilable with some observational bounds\footnote{ Such a
running has not been observed yet, but Planck provides the following
bound (again from the combined data from temperature fluctuations
and lensing \cite{Ade:2015xua,Ade:2015lrj}; $\alpha_s=-0.003\pm
0.007$,~~~(68\% CL)}.

However, future observations may allow one to discriminate between
models (inflationary, Ekpyrotic and matter bounce scenarios), in
this time, we are interested to introduce some deformed models of
matter bounce to obtain a negative running $\alpha_s$, like the
inflationary scenario \cite{Lehners:2015mra}.

After introducing $\Lambda$CDM matter bounce scenario by Cai et al.
\cite{Cai:2014jla}, new insights into the deformed matter bounce
scenario is provided. In this paper authors considered a
cosmological constant (vacuum energy) as a dark energy term with a
constant equation of state (EoS) parameter ($w_{\Lambda}=-1$,
accompanied with a pressureless cold dark matter (CDM)). The
effective EoS parameter does not remain constant (slightly
increasingly negative) in this setting and it eventually provides a
slight red tilt in spectral index (an small value less than unity of
spectral index $n_s$), according to observations. Finally, they
obtained a positive value for the running of scalar spectral index.
Another model of deformed matter bounce with dark energy introduced
by Odintsov et al. in order to describe the late time acceleration
of the universe \cite{Odintsov:2016tar}. In this model authors
considered a deformation that affects the cosmological evolution,
only at the late time not at beginning of contraction phase. They
showed that the big rip singularity can also be avoided in their
model. These models give us a great motivation to consider some
model of matter bounce with various forms of dark energy to solve
another remained problem.

Recently running vacuum models of dark energy (RVM-DE) on the basis
of renormalization group \cite{Basilakos:2012ra} has been attracted
a great deal of attention \cite{Sola:2016vis,Sola:2016jky,
Fritzsch:2016ewd,Sola:2016ecz,Sola:2017jbl,Perico:2013mna}. In these
models, not only the vacuum energy $\Lambda$ has been considered as
a series of powers of Hubble rate $H$ and its first time derivative
\cite{Shapiro:2000dz,Sola:2016jky}, but also it gets a constant
equation of state parameter same as the cosmological constant. Then
the energy density of RVM-DE reads:
\begin{equation}
\rho_{\Lambda}(H)=\alpha_0+\sum_{n=1} (\alpha_n H^n+\beta_n\dot
{H}^n).
\end{equation}
In standard cosmology, these models strongly preferred as compared
to the conventional rigid $\Lambda$ picture of the cosmic evolution
\cite{Sola:2016jky}. Due to these evidences, studying on a modified
matter bounce scenario with a class of RVM-DE attracts a great deal
of attention.

This paper is organized as follows: In Sec. \ref{sec2}, we give a
brief review on bouncing cosmology with a dynamical vacuum energy.
Then, as a simple example we study on the standard $\Lambda$CDM
cosmology in bouncing scenario in Sec. \ref{sec3}. We extend this
model with RVM-DE in sec \ref{sec4}. In Sec. \ref{sec5}, the study
of cosmological perturbation theory takes placed analytically for a
simple case. The spectral index and its running are calculated
numerically for some other cases of (RVM) model in Sec. \ref{sec6}
and at last we finished our paper by some concluding and remarks.

Before getting started, it must be noted that we are using the
reduced Planck mass unit system in which $\hbar=c=8\pi G=1$ and also
considering a flat Friedmann-Lema\^{i}ter-Robertson-Walker (FLRW)
metric, with the following line element
\begin{equation}
ds^2=-dt^2+a(t)^2\sum_{i=1,2,3}(dx^i)^2.
\end{equation}

\section{Cosmological bounce with dynamical vacuum energy } \label{sec2}

 First we give a brief review on the dynamics of bouncing cosmology in a flat FLRW universe with time varying $\Lambda (t)$ model.
 The matter contents are composed of radiation and cold dark matter (CDM).

In high energy cosmology, a Holonomy corrected Loop Quantum
Cosmology (LQC) gives approximately full quantum dynamics of the
universe by introducing a set of effective equations
\cite{Ashtekar:2006wn} \barr  H^2&=&\frac
{\rho_{tot}}{3}(1-\frac{\rho_{tot}}{\rho_c}), \label{3}\\ \dot
H&=&(\frac{1}{2} \rot-3H^2)(1+w), \label{5}
 \earr
 where
$ \rho_{tot}= \rho_m+\rho_r+\rho_{\Lambda} $ is total energy density
of pressureless CDM, radiation and dark energy respectively. The
quantity $\rho_c$ is the critical energy density. In fact the
magnitude of this parameter is model dependent, namely, the
contribution of the corrections arisen from the specific Holonomy
forms. Nevertheless, the upper bound of this parameter is less than
the Planck density. From Eqs. \eqref{3} and \eqref{5}, the
continuity equation of total energy density easily obtained
 \bq
 \dot{\rho}_{tot}+3H\rho_{tot}(1+w)=0, \label{4}
\eq
  where $w$ is the  effective equation of state parameter
  $\Pot=w\rot$. The continuity equation (\ref{4}) can be decomposed
  by the following equations for all components of energy as
 \barr
   \dot{\rho}_m+3H\rho_m=& \dot{\rho_{\Lambda}}, \\
 \dot\rho _r+4H\rho_r=& 0 ,\label{1}
  \earr
 where the superscript dot refers to derivative with respect to cosmic
 time. This model generally named dynamical vacuum model (DVM) in
 which the EOS parameter is still $w_D=-1$, like as a rigid $\Lambda$ model \cite{Perico:2013mna}.
 Note that if $ \rho_c \to \infty $, the classical Friedmann equations in the flat universe are retrieved.
 Now assume that the bounce occurs at $t=0$, so that at this time we have $H=0$ and $\rot(t=0)= \rho_c\approx
 \rho_r$. In fact around the bounce point, radiation is considered as the dominant term of energy
 density.

 In terms of conformal time $\eta$\ in which $d \eta = {dt}/{a} $, all previous effective equations can
be rewritten as
\barr \rho_{tot}^\prime&=& -3\mH\rho_{tot}  (1+w), \label{c1}\\
   \mH^2&=& \frac {\rho_{tot}}{3} a^2 (1-\frac{\rho_{tot}}{\rho_c}),
   \label{confbounce}\\
 \mH^\prime&=&\frac{a^2}{2} \rho_{tot} (1+w)-\mH^2(2+3w), \label{background3}
\earr
 where $\mH ={a^\prime}/a=a H$ is the Hubble rate in conformal time and prime denotes derivative with respect to conformal time
$\eta$. Also for convenience, the scale factor can be normalized to
unity at the bounce point ($a(\eta=0)=1$).

   \section{Bouncing with the standard $\Lambda$CDM}\label{sec3}
In this model we are using the vacuum energy as a dark energy
$\rho_{\Lambda}=\Lambda$. By solving effective Eqs. \eqref{c1},
\eqref{confbounce} and \eqref{background3}, we can find the
evolution of cosmological parameters. Although from LQC the value of
$\rho_c$ is roughly equal to the Planck energy density, observed
amplitude of scalar perturbations
  in matter bounce scenario required $ \rho_c \thicksim 10^{-9} \rho_{pl}  $ \cite{WilsonEwing:2012pu}.
  It means that the bounce occurs at much lower energy in this scenario.
  The continuity equation \eqref{1} for matter yields
  \bq
    {\rho}^\prime_m+3\mH\rho_m =0,
  \eq
  and consequently the total energy density becomes
  \bq
    {\rho}_{tot}=\rho_{im}\left(\frac{a_i}{a}\right)^3+\rho_{ir}\left(\frac{a_i}{a}\right)^4+\rho_{\Lambda}, \label{13}
  \eq
  where subscript '$i$' refers to initial condition. Taking critical
  energy density at bounce point and initial conditions in reduced Planck
  mass unit, same as \cite{Cai:2014jla}, as follow
  $$\rho_c=2.9 \times 10^{-9},   $$
  $$\rho_{im}=1.1 \times 10^{-24},$$
  $$ \rho_{ir}=5.1 \times 10^{-28}.$$
   These values are selected in a way that quite far from the bounce
   point in contracting matter dominated universe, the ratio of $\rho_{im}$ to $\rho_{ir}$ is nearly the same as present
   time in standard cosmology. Also matter-energy density has 15 orders of
   magnitude less than the energy density of bounce point and again we emphasize that
   around the bounce point it usually considered $\rho_c\simeq \rho_r$.

\begin{figure}[!h]

        \begin{center}
        \includegraphics[scale=0.42]{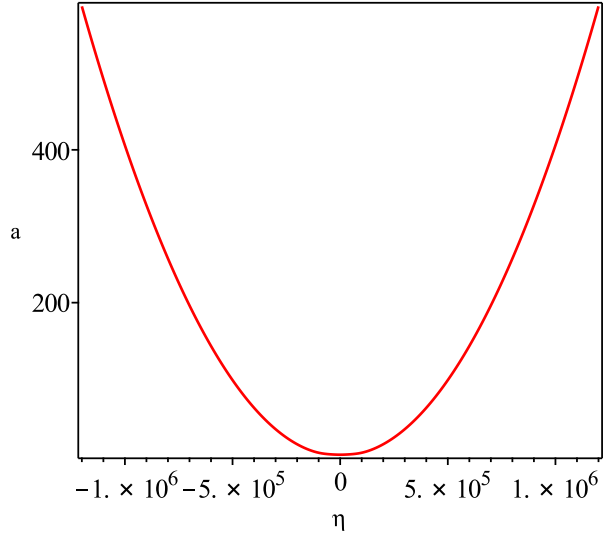}
        \caption{\footnotesize The evolution of the scale factor versus $\eta$. At bounce point ($\eta=0$), it gives a positive non-zero value}
         \label{fig4}
          \end{center}
          \end{figure}

 Using above initial conditions and effective equations (\ref{c1}, \ref{confbounce},
 \ref{background3}), the evolution of the scale factor versus to
 conformal time $\eta$ is obtained by numerical computation. In fig.
 \ref{fig4}, we see that in this case our universe is expanding after
 a contracting phase through a non-singular point (big bounce).

    \section{Bouncing with Running vacuum model} \label{sec4}
   The running vacuum energy in quantum field theory (QFT) in curved space-time motivated us to consider
    $\rho_{\Lambda}=\Lambda(H)$ in reduced Planck mass unit. This theory gives the renormalization group
    equation \cite{Basilakos:2012ra}

           \bq
           \frac{d \rho_{\Lambda}}{d \ln(\mu^2)}=\f{1}{4\pi}\left( \sum_{i}{B_i M_i \mu^2}+\sum_i{C_i \mu^4}+...  \right), \label{QFT}
           \eq
            where $\mu^2$ can be a linear combination of $H^2$ and $\dot{H}$ \cite{Sola:2015rra}, $B_i$ and $C_i$
            are dimensionless coefficients and $M_i$ is the mass of any particle which
            contribute in the dynamics.

             By setting $\mu^2=H^2$, the equation \eqref{QFT} simply yields

           \bq
           \rho_{\Lambda}(H^2)=n_0 + n_2 H^2 +n_4 H^4 +O(H^6).\label{H^2n}
           \eq
         However in general, by consideration of correction of QFT, a theoretical
         explanation of RVM-DE becomes (\cite{Sola:2013gha,Sola:2015rra,Sola:2016jky} and reference therein)

   \bq
   \rho_{\Lambda}(H^2,\dot H)=n_0+n_2 H^2+\beta\dot H+n_4 H^4 +O(H^6).\label{varlambda}
   \eq

    The terms with higher powers of Hubble function have recently been used to
    describe inflation \cite{Sola:2013gha,Basilakos:2013xpa,Basilakos:2014moa,Lima:2015mca,Sola:2015csa}.
    For $n_i=\beta=0;~~i=1,2,3,...$, the standard $\Lambda$CDM is recovered. On the other
    hand any model with a linear term of Hubble function motivated from a phenomenological point of view
    \cite{Schutzhold:2002pr,Thomas:2009uh,Urban:2009vy,Urban:2009ke,Urban:2009wb,Klinkhamer:2009ri}.
    This model can still be tenable if a constant additive term
    is around \cite{Gomez-Valent:2014fda}. It must be mentioned that a model of vacuum
    energy in which the density is just proportional to $H$ can not
    come from any covariant QFT and it does not even have a well-defined
    $\Lambda$CDM limit and the worst, it is also excluded from the data
    on structure formation \cite{Gomez-Valent:2014fda}.

In following we are also interested to add another term $n_1 H$ into
Eq. (\ref{varlambda}) and studying on the role of each terms on
evolution of the scale factor, density parameters,
    Hubble parameter and equation of state parameter in some cases and compare them with the $\Lambda$CDM bouncing model.

        \section{cosmological Perturbation theory } \label{sec5}
         The dynamics of scalar perturbations on
         a spatially flat background spacetime are explained by the Mukhanov-Sasaki equation with a
         gauge invariant variable $v=z \mR$ \cite{Mukhanov:1990me} in which $\mR$ is the comoving curvature perturbation and
          \bq
          z = \f{a \, \sqrt{\rho_{tot}+P_{tot}}}{c_s \, H}=\f{a \, (1+w)^{1/2} \, }{c_s }\frac{\sqrt{\rho_{tot}}}{H}\label{z}.
          \eq
         Linear perturbations can be extended into LQC \cite{Salopek:1990jq, Wands:2000dp}.
         The LQC effective equation for the Mukhanov-Sasaki variable is
          \cite{WilsonEwing:2011es, WilsonEwing:2012bx}

            \bq \label{pertur}
               v'' - c_s^2 \, \left(1 - \f{2 \rho_{tot}}{\rho_c} \right)
              \nabla^2 v - \f{z''}{z} v = 0,
            \eq
          where $c_s$ is the speed of sound which is a constant parameter depending on every epoch of history of the
          universe. The effective equation \eqref{pertur} is expected to provide a good approximation to the
          full quantum dynamics for modes that always remain large compared to the Planck length
          \cite{Rovelli:2013zaa}.
          It must be noted that for $\rho_c \to \infty$, the standard classical perturbation equation is recovered \cite{Mukhanov:2005sc}.
           Also the Holonomy-corrected tensor perturbation in LQC is \cite{Cailleteau:2012fy}:

    \bq \label{perturtensor}
    \mu'' -  \, \left(1 - \f{2 \rho_{tot}}{\rho_c} \right)
     \nabla^2 \mu - \f{z_T''}{z_T} \mu = 0,
    \eq

    where $\mu= h \, z_T $  in which
    \bq
    z_T=\frac{a}{\sqrt{1-2\rho/\rho_c}}.
    \eq

\subsection{Analytical Solutions with case $\rho_{\Lambda}=n_2
H^2+\beta\dot{H}$}

Following \cite{Cai:2014jla}, we consider three continues era for
studying on  scalar perturbation and  power spectrum in analytical
method. Modes of interest are those that reach the long wave length
limit during the first era where it is very far from the bounce in
which the evolution of the contracting universe treats as
matter-dark energy domination. Then after equality of radiation with
previous pair components, the universe enters to a radiation
domination epoch and at last, goes through the bounce where the
evolution of the universe governed by LQC.

Far enough the bounce, where quantum gravity effects are negligible,
effective equations are standard Friedmann equations
\begin{align}
      &3 H^2=\rho, \nonumber \\
       &\dot{\rho}=-3 H \rho (1+w)\label{fridmann},\\
       &\dot{H}=\dfrac{-3}{2} H^2(1+w) \label{hdot},
       \end{align}
       where we are using $\rho$ instead of $\rho_{tot}$ for simplicity.
From above equations, the EoS parameter can be found simply as
         \bq
        w= w_0=-\frac{2 n_2-3\beta}{6-3\beta}. \label{w0}
         \eq
It is important to note that in a matter-dark energy dominated
epoch, in order to have a nearly scale invariant power spectrum, the
effective EoS parameter $w_0$ must be slightly negative.

 Fortunately, we can directly calculate power spectrum and  spectral index same as
 Mukhanov's method of inflationary scenario, in conformal time \cite{Mukhanov:2005sc}.
 Very far from the bounce, using \eqref{fridmann}, the $z$ quantity in
 (\ref{z}) reduced to
   \bq
   z=a \sqrt{3}\left(\frac{\sqrt{1+w_0}}{c_s}\right),
   \eq

   and the Mukhanov-Sasaki equation for scalar perturbations (\ref{pertur}) in Fourier
   modes will be rewritten as

   \bq
    v'' + (c_s^2 \, k^2  - \f{a''}{a} )\, v = 0. \label{pertur21}
   \eq

  Solving second Friedmann equation \eqref{hdot} in conformal time

    \bq
    \mH ' =- \f{\mH^2}{2}(1+3w_0), \label{confH}
    \eq
yields

     \bq
   \mH= \frac{2}{(1+3 w_0)(\eta-\eta_0)},
   \eq
where $\eta_0$ is the constant of integration.
   Taking $a''/a = \mH'+\mH^2$ and for very small values of $w_0$, the Mukhanove-Sasaki equation (\ref{pertur21})
    approximately yields

      \bq
      v''+\left(c_s^2 k^2 -\dfrac{\nu^2 -\dfrac{1}{4}}{(\eta-\eta_0)^2}
      \right) v=0,  \label{pertur2}
      \eq
     where $\nu=(9/4-18 w_0)^{1/2}\approx 3/2-6 w_0+O(w_0^2)$. The relevant solution is
       \bq
       v_k= \sqrt{|\eta-\eta_0|}\left(C_1 \, H^{(1)}_\nu \left(c_s k |\eta-\eta_0| \,\right) +
       C_2 \, H^{(2)}_\nu \left(c_s k |\eta-\eta_0| \,\right) \right). \label{answer}
       \eq
   Assuming the initial conditions of primordial perturbations in the distant past of the pre-bounce epoch, to be quantum vacuum
   states, it takes
      \bq
      v_k=\frac{{1}}{\sqrt{2 c_s k}}e^{-i \, c_s k \
      \eta}.
      \eq
       Using the asymptotic behavior of the first type of the Hankel function when $|\eta-\eta_0|>>0$
       \bq
        H^{(1)}_\nu \left(c_s k |\eta-\eta_0| \,\right)= \frac{\sqrt{2} \,
          e^{-(\nu+\frac{1}{2})\f{\pi}{2}}}{\sqrt{c_s k  \pi |\eta -\eta_0|}}\,e^{i \, c_s k |\eta-\eta_0|
          },
       \eq
in solution (\ref{answer}), we find
 \bq C_1=\frac{\sqrt{\pi}}{2}
e^{i(\nu+\frac{1}{2})\f{\pi}{2}};~~~C_2=0.
 \eq
 In long wavelength limit, $ c_s k |\eta - \eta_0|<<1$ and small values $|w_0|$, the solution reduced to
\bq
  v_k\cong \f{- i}{2} \sqrt{\f{1}{2}}
  \,(c_s k)^{-3/2+6w_0} \left(\mH \right)^{1-6w_0}.
 \label{finall}
\eq

In second step, after equality, before the quantum gravity effects
become considerable, the evolution of the universe tends to
radiation dominated epoch in which $w=1/3$, $c_s=\sqrt{1/3}$ and
$\rho_{tot}=\rho_{0r}/a^4$. Also Eq. (\ref{confbounce}) in the limit
$\rho_c\rightarrow \infty $, gives
   \begin{equation}
    a(\eta)=\sqrt{\frac{\rho_{0r}}{3}} \eta;~~~~~
    \mH =\frac{1}{\eta}. \label{conftot}
  \end{equation}
 The perturbation equation reduces to a
harmonic oscillator
  \begin{equation}
    v'' + c_s^2 k^2 v = 0, \label{osil}
  \end{equation}
  with the following solution
 \begin{equation}
 v_k=B_1 \sin  (\frac{k}{\sqrt{3}} \eta)+B_2
\cos(\frac{k}{\sqrt{3}} \eta).
\end{equation}
Since this step is between dark matter-energy domination era and
bounce period, hence in order to  continue with the bounce period
($\eta \rightarrow 0$), the coefficient $B_2$ will dominate and
hence it requires $ B_1=0$. On the other hand the continuity of $v_k
$ and $v_k'$ at equality time $\eta_e$, gives
   \bq
   B_2=-\f{\sqrt{3}}{k} \sin (\frac{k}{\sqrt{3}} \eta_e) v_k'|_{\eta=\eta_e}+\cos(\frac{k}{\sqrt{3}} \eta_e) v_k |_{\eta=\eta_e}, \label{B22}
   \eq
   and after substituting \eqref{finall} and its derivative into \eqref{B22}
   it goes
    \begin{align}
    B_2=&-\f{i \sqrt{3} (c_s k)^{-3/2+6w_0}}{4\sqrt{2 }k} \sin (\frac{k}{\sqrt{3}} \eta_e) \mH_e^{2-6w_0}\nonumber
     \\ &-\f{i  (c_s k)^{-3/2+6w_0}}{2\sqrt{2 }} \cos (\frac{k}{\sqrt{3}} \eta_e)
     \mH_e^{1-6w_0}.
    \end{align}

    An essential condition  for  scale invariant power spectrum  is $k |\eta_e| <<1$ \cite{Cai:2014jla}.
    Therefor, in this approximation,  $B_2$ simplify to

    \bq
    B_2 \cong -\f{3 i}{4}  \sqrt{\f{1}{2}} (c_s k)^{-3/2+6w_0}  \mH_e^{1-6w_0}. \label{apB2}
    \eq

      At last, in the bounce period, since  the  radiation is yet a dominant term ($w=1/3$), the first LQC effective equation in conformal time
      gives

      \begin{equation}
      \mH^2=\f{\rho_{0r}}{3 a^2}(1-\f{\rho_{0r}}{\rho_c a^4}),
      \end{equation}
      which directly gives the following scale factor
      \bq
       a=(a_0 t^2+1)^{1/4}, \label{at}
       \eq
       where $a_0= {4 \rho_c}/3$. On the other hand, from \eqref{pertur}, the perturbation equation in conformal time becomes

        \begin{equation}
        \frac{v''}{v}-\frac{z''}{z}=0,
        \end{equation}
        with following solution

        \begin{equation}
        v(t)=\f{C_1 a_0}{4 \sqrt{\rho_c}} \, z \, \,t^3  F \left(\f{3}{2},\, \f{7}{4} ,\, \f{5}{2},\, -a_0 t^2\right)+ \f{a_0 z}{4
        \sqrt{\rho_c}},
       \end{equation}
     after using \eqref{at}. Its asymptotic behavior governs
       \bq
\mathcal{R}\cong \f{v_k}{z}=\f{2 A_1 a_0^{-3/4} \sqrt{2\pi^3}} {z \Gamma(\f{3}{4})^2}=\f {1}{6} \sqrt{\f{\pi}{3}}
 \f{\Gamma( \f{1}{4})}{\Gamma(\f{3}{4})} B_2,
\eq
where
      \bq
       A_1=-\f{a_0 B2}{6}.
       \eq
 Thus the power spectrum for modes that become (nearly) scale invariant in which $k|\eta_e| << 1$,
 is given by

       \bq
              \Delta^2=\f{k^3 \,| \mathcal{R}|^2}{2 \pi^2} = \f{1}{768} \left( \f{\Gamma( \f{1}{4})}
              {\Gamma(\f{3}{4})}\right)^2 c_s^{-3 +12 w_0} \mH_e^{2-12 w_0} \, k^{12 w_0}.
        \eq
It is worthwhile to mention that these modes must also remain
outside the sound Hubble radius during the entire the contracting
radiation dominated epoch and bounce period as well as dark
matter-energy domination contracting era.

       At the end, from \eqref{conftot}, power spectrum can be
       rewritten as

       \bq
        \Delta^2=\f{k^3 \,| \mathcal{R}|^2}{2 \pi^2} = \f{ 1}{768} \left( \f{\Gamma( \f{1}{4})}
        {\Gamma(\f{3}{4})}\right)^2 c_s^{-3 +12 w_0} \left(\left(\f{ \rho_c}{3}\right)^{1/4}
        \sqrt{H_e}\right)^{2-12 w_0} \, k^{12 w_0}.
       \eq
  Obviously, if we set $w_0=0$, the power spectrum will be exactly scale-invariant, but for a small negative value of $w_0$, it
  is nearly scale-invariant with a red tilt for spectral index as it is predicted by observational data ($n_s=1+12 w_0$)

In this section we study the Fourier modes evolution which they come
from initial quantum vacuum state far away the bounce. Any mode that
exit the sound Hubble radius in matter-dark domination period
becomes scale invariant and they return to sound Hubble radius after
the bounce, in expanding branch.

     \section{Spectral Index and its Running}\label{sec6}

     Considering the semi-matter dominated epoch in contracting phase of the deformed matter bounce scenario. Far from the bounce point,
   the EoS parameter becomes very small $(w<<1)$ and approximately constant. Its time derivative is
   very small $ (w'<<1)$ around the crossing time, when the sound horizon crossed by long wavelength
  modes (e.g. see Fig. 3). It gives a good condition to solve the perturbation equation
   analytically and consequently obtains a simple relation for the power spectrum index (same as the previous section).
Expanding the parameter $w$ around $\eta_0$ up to first order \bq
w\approx w_0 + \kappa (\eta-\eta_0), \label{tayexp} \eq where
$\kappa=d w/d \eta$ at $\eta=\eta_0$ and $w_0$ is the value of $w$
at $\eta_0$. Since the changes of $w$ is very small, $|\Delta w|
\approx |w-w_0||\kappa (\eta-\eta_0)|<<1$, therefore during the
semi-matter dominated epoch in a contraction phase, at low curvature
and energies, in Eq. \eqref{pertur} we have $\frac{z''}{z}\cong
\frac{a''}{a}$ (details are referred to \cite{Elizalde:2014uba}).

Now by this approximation, the conformal Hubble parameter and
$a''/a$ are calculated as \bq \mH=\frac{4}{3
\kappa(\eta-\eta_0)^2+2(3w_0+1)(\eta-\eta_0)}, \label{mH1} \eq \bq
\frac{a''}{a}\cong\frac{2-18 w_0}{(\eta-\eta_0)^2}-\frac{12
\kappa}{\eta-\eta_0}. \eq

By setting $\nu\cong 3/2 -18 w_0$ and $\xi=12 \kappa$, the scalar
perturbation equation \eqref{pertur} in Fourier mode becomes

\bq
 v''+c_s^2 k^2 \left(1 -\dfrac{\nu^2 -\dfrac{1}{4}}{c_s^2 k^2 (\eta-\eta_0)^2}+\frac{\xi(\eta-\eta_0)}{c_s^2 k^2
(\eta-\eta_0)^2} \right) v=0.  \label{prt} \eq It is worthwhile to
mention that in limiting case, when $|\xi (\eta-\eta_0)|\rightarrow
0$, the above equation reduced to Eq. \eqref{pertur2} in the
previous section.

For convenience in calculations, we replace $\eta-\eta_0$ in
\eqref{prt} with $\eta$, \bq
 v''+ \left(c_s^2 k^2 -\frac{\nu^2 -\dfrac{1}{4}}{\eta^2}+\frac{\xi}{\eta} \right) v=0.
\eq General answerer of this equation is \bq v\,= J_1 \, W\left(
\frac{-i \, \xi}{2 \, c_s \, k},\,\nu\, , 2 \, i \, c_s \, k \,
\eta\right) +J_2 \, M\left( \frac{-i \, \xi}{2 \, c_s \, k},\,\nu\,
, 2 \, i \, c_s \, k \, \eta\right), \label{witsol}\eq where  $W$
and $M$ are Whittaker functions and $J_i$'s are constants of
integration. By considering the asymptotic behavior of $W$ at large
$|k \eta|$,
 \bq
  W\left( \frac{-i \, \xi}{2 \, c_s \, k},\,\nu\, , 2 \, i \, c_s \, k \, \eta\right)  \sim\exp(-i\, c_s \, k \eta) (-2 i c_s\, k \eta)^{-i \xi/(2c_s
 k)},
\eq and taking the initial condition of primordial perturbations to
be quantum vacuum states, we find $J_2=0$ and
 $$J_1= \sqrt{\frac{1}{2 c_s k}} \exp(\frac{\pi \xi }{4 c_s k}). $$

Now in the long wavelength limit, around the crossing horizon where
$(|c_s k \eta| << 1)$, the solution of \eqref{witsol} rewritten as
\begin{align}
v &= \sqrt{\frac{1}{2 c_s k}} \, \,  \exp(\frac{ \pi \xi}{4 c_s k})
\frac{\Gamma(2 \nu )}{\Gamma\left(1/2+ \nu+\dfrac{i \xi}{2 c_s k}
\right)}\, (2\, i \, c_s k \, \eta)^{1/2-\nu} \nonumber\\ & \cong
\frac{-i}{2}\sqrt{\frac{1}{2 c_s k}} \exp(\frac{\pi \xi }{4 c_s k})
\, \, (\frac{1}{2}c_s \, k \, \eta)^{-1+6 w_0},
\end{align}
which it will reduce to Eq. \eqref{finall} limiting case
$\xi\rightarrow 0$.

Now, completely similar to the previous section, since the
 perturbations must be continued during the contraction
and expansion of the universe, after forward calculation, the power
spectrum $\Delta^2$ will be modified by coefficient $C(k)=\exp[\pi
\xi / (2 c_s k)] $ as follows \bq \Delta^2(k)\sim
\exp(\frac{\pi\xi}{2 c_s k}) \,\, k^{12 w_0}. \eq

The spectral index in this case becomes \bq n_s-1=\frac{\di ln
\Delta^2}{\di ln k}= -\frac{\pi \xi }{2 c_s k}+ 12 w_0.
\label{newspect}\eq Also to obtain the running of spectral index, as
we will also be pointed out later, \bq \alpha_s=\frac{\di n_s}{\di
\ln k}= \frac{\pi \xi }{2 c_s k}=\frac{6 \pi \kappa }{ c_s k}. \eq
An interesting point of this relation is obviously if $\kappa=d w/d
\eta<0 $, at any time, especially at crossing time, this running
becomes negative. This point will be hinted again at next sections.
Another point is about the value of the running of spectral index.
From the observational data, $\alpha_s$ is very small and negative.
Therefore, it is required that in the long wavelength limit, \bq
|\frac{ \pi \xi }{2 c_s k}|\ll 1 \Rightarrow |\xi (\eta-\eta_0)| \ll
|c_s k (\eta-\eta_0)|\ll 1, \eq which again it emphasizes that the
second term in Eq. \eqref{tayexp} is very small in a semi-matter
bounce scenario.

Finally, the Eq. \eqref{newspect} at the crossing Hubble horizon,
will be rewritten as \bq
n_s=1+12(w_0+\frac{|\alpha_{sc}|}{12})\approx 1+12 w,
\label{spectralxi}\eq

where $\alpha_{sc}$ is the running of spectral index at the crossing
Hubble horizon which is a very small negative value, approximately
same as the role of $|\kappa (\eta-\eta_0)|$ in Eq. \eqref{tayexp}.
Also Eq. \eqref{spectralxi} will reduce to $n_s=1+12 w_0$ for
constant EoS parameter of DE-model and vanishing $\alpha_s$, same as
the previous section. It is worthwhile to mention again that we
concentrate our attention to semi-matter bounce scenario which has a
very small values of EoS parameter at whole of the contracting
phase.

Moreover with this assumption, when a mode is crossed by Hubble
radius, by using \eqref{spectralxi},
 the running of spectral index yields (details are referred to
 \cite{deHaro:2015wda}),

     \bq
   \alpha_{sc}= \left( \frac{ d n_s}{ d \ln k}\right)_{k=a|H|
   }= \left(\frac{ d n_s}{ d t}  \frac{ d t}{ d \ln k} \right)_{k=a|H|}=\frac{12H \dot{w}}{H^2+\dot{H}} .\label{alphass}
     \eq
    In a RVM model, in which $ P_D=-\rho_D $, far enough the bounce, from the first Friedmann equation,
    the effective EoS parameter gives

     \bq
     w=- \frac{\rho_{\Lambda}}{\rho_{\Lambda}+\rho_{m}}=-
     \frac{\rho_{\Lambda}}{3H^2}\label{ww},
     \eq
     and after some calculations, $\dot{w}$ becomes
      \bq
      \dot{w}=\frac{\dot{\rho_\Lambda}}{\rho_\Lambda}w+3
      Hw(1+w).\label{dw}
      \eq
     Using Eqs. \eqref{ww} and \eqref{hdot}, the Eq. \eqref{dw} will be rewritten by
     \bq
     \dot{w}=\frac{H^2(w+1)}{2}\frac{d}{d H}\left(\frac{\rho_{\Lambda}}{H^2}\right). \label{wdot}
     \eq
     According to $ |w|<<1$, $w$ is neglected and $\alpha_s$ becomes

     \bq
     \alpha_s\cong-12
     H  \frac{d}{d H}\left(\frac{\rho_{\Lambda}}{H^2}\right), \label{running}
     \eq
     so, approximately, the relation \eqref{running} can be rewritten by the following simple form
     \bq
     \alpha_s\cong3 H \frac{d n_s}{d H}. \label{simplerunning}
     \eq

       In agreeing with the $\Lambda$CDM bouncing model\cite{Cai:2016hea}, if
       $\rho_{\Lambda}=n_0$, from \eqref{running}, we can see that the running of the spectral
       index becomes positive which is an obvious weakness of this case.
          Note that the effective equation of state parameter is negative in contracting phase of the universe at crossing
          time. In this case $\alpha_s$ approximately given by
          \bq
          \alpha_s=24 \frac{n_0}{H^2}=-72 w,
          \eq
          which gets a positive value for $w<0$.

          For a constant effective EoS parameter, same as previous case
          (Eq. \eqref{w0}), running of spectral index is vanishing which is in
          contrast with Planck bound \cite{Ade:2015xua,Ade:2015lrj}. It is worthwhile to mention that also in the
          standard cosmology this type of RVM-DE (model of Sec. \ref{sec5}) has been already
          excluded on account of its inability to correct
          description of the data on structure formation
          \cite{Gomez-Valent:2014rxa,Gomez-Valent:2014fda,Gomez-Valent:2015pia}.
          Thus this fact give us an alternative reason to exclude this type of RVM-DE.

          In order to have a negative value of running of spectral index, ($\alpha_s<0$), which
          is compatible with the inflationary paradigm, it is required that $H \frac{d}{d H}
          (\rho_{\Lambda}/H^2)>0$ (see Eq. \eqref{running}).

At following we will give two other cases of RVM and will calculate
the spectral index and running.

      \subsection{ Case $\rho_{\Lambda}=n_0+n_2H^2$}
      This is one of the known cases of RVM which has been studied by many authors in
      standard cosmology
      \cite{Sola:2007sv,Basilakos:2009wi,EspanaBonet:2003vk,Shapiro:2004is,Basilakos:2012ra}.
      In bouncing scenario, according to previous section, far enough the bounce point,
      when the sound horizon crossed by long wavelength modes, from  \eqref{spectralxi}  and \eqref{running}, we will have

     \bq
     n_s -1 =-12( \frac{n_0}{3 H_{cr}^2} + \frac{n_2}{3}),
     \eq
and the running will be
     \bq
     \alpha_s=24 \frac{n_0}{H_{cr}^2},
     \eq
      where $H_{cr}$ is the value of the Hubble parameter at crossing time ($\eta=\eta_{c}$).
As a result, $\alpha_s$ is always positive unless $n_0<0$. However a
negative value of $n_0$ is forbidden near the bounce point where
$H\approx 0$ and consequently $\rho_{\Lambda}\approx n_0$.

 \subsection{$\rho_{\Lambda}=n_0 +  n_2 H^2+n_4 H^4$}
The first attempts to consider this type of RVM which was extended
to $H^4$ term in standard cosmology was given by in
\cite{Lima:2012mu}.
 In this case the effective EoS parameter is simply calculated as
          \bq
          w_0=\f{n_s-1}{12}=-\frac{1}{3}\left(\f{n_0}{H_{cr}^2} +{n_2}{}+{n_4}{} H_{cr}^2\right),
          \eq
           and from Eq. \eqref{running} the running $\alpha_s$ is

           \bq
           \alpha_{sc}=24 \left(\f{n_0}{H_{cr}^2} - n_4
           H_{cr}^2\right).\label{runningcase2}
           \eq
           Obviously in order to have $\alpha_{sc}<0$, it requires that
           \bq
           n_4 > \f{n_0}{H_{cr}^4}. \label{ineq}
           \eq
           Simply after some algebraic calculation, the relation between $n_0$, $n_2$ and $n_4$, can be
           found as

           \bq
           n_2=\f{6-6n_s- \alpha_{sc}}{24}-\f{2n_0}{H_{cr}^2}, \label{n22}
           \eq

           \bq
           n_4=\f{n_0}{H_{cr}^4}-\f{\alpha_{sc}}{24 H_{cr}^2}.  \label{n44}
           \eq

 These relations can help us to estimate the order
 of magnitude of parameters of this case to find benefit numerical calculations at next.

   It is important to note that in cosmological perturbation
      theory, the power spectrum, spectral index and its running essentially
      depend on the effective equation of state and its derivative at
       time of horizon-crossing. In other words, in the contracting phase, the space-time curvature is not felt by the Fourier
      modes inside the horizon, so they  oscillate until  exiting inside the horizon.
      The background spacetime evolution, equation of state and the time derivative of $w$ at the horizon-crossing
      in matter dominate epoch have an important role to appropriate predict of nearly scale-invariant
      power spectrum  and the running of the spectral index.
      Now we ask, is it possible to have a negative running of the spectral index in matter bounce scenario in this case?

      To answer this question, firstly, we interested to solve numerically the background  differential equations
      (\ref{c1}, \ref{confbounce}, \ref{background3}).

         \subsubsection{background numerical calculation  }

         By using Eqs. \eqref{n22}, \eqref{n44}, and the requirement of negative running in \eqref{runningcase2},
         a set of parameters can be estimated as
    \bq
    n_0=2.2 \times 10^{-28} \,\,\,\,\,\,\, n_2=8.9 \times 10^{-3} \,\,\,\,\,\,\, n_4=4.47 \times
    10^{10} \, . \label{parameters}
    \eq
   It is worthwhile to mention that in reduced Planck mass
    units, $\rho_{\Lambda}$ has dimension $H^2$ and $\dot{H}$, i.e. inverse length squared,
    (see Eqs. (\ref{3}, \ref{5})) and consequently time, length and mass get equal dimension.
    Therefore in this case, the constant parameter $n_0$ has dimension $H^2$, the coefficient of $H^2$, $n_2$, is dimensionless
     and the coefficient of $H^4$, $n_4$, has dimension $H^{-2}$ (length squared).
     Also, we must note that this set of parameters achieved approximately
by a rough estimating of constrained values of $\alpha_{sc}$, $n_s$
and the calculated value of $H_{cr}$ at $\eta_{c}$. These
coefficients satisfy inequality \eqref{ineq} and reminded that in
bounce point
   the universe is dominated by radiation.
    The evolution of the scale factor, as expected, similar to
    fig.\ref{fig4}, shows an expansion after contraction through the critical bounce point without any singularity.

\begin{figure}[!]
      \begin{center}
      \includegraphics[scale=0.42]{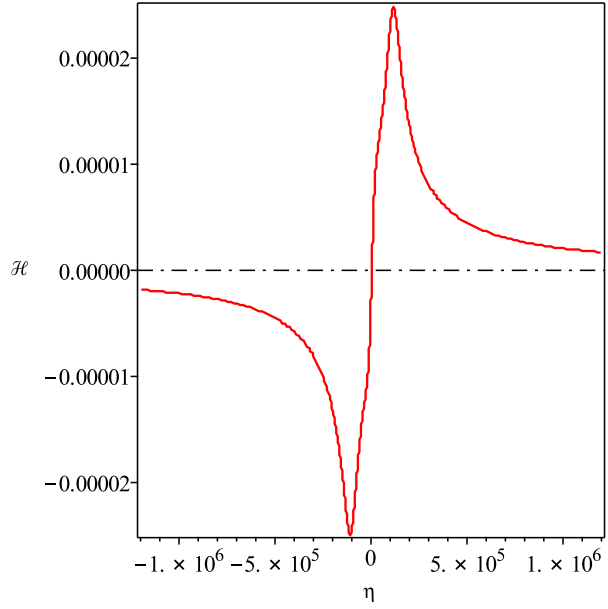}
      \caption{\footnotesize The evolution of the conformal Hubble parameter  $\mH$ versus $\eta$. At bounce point ($\eta=0;~~\mH =0$),
       it shows a transition from a contracting universe to an expanding one from left to right.}
       \label{conformalH}
        \end{center}
        \end{figure}

    The evolution of the conformal Hubble rate is indicated in fig. \ref{conformalH}.
    The value of $\mH$ is decreasing into a minimum value when $\eta\approx -1\times10^5$. After that, $\mH$
    increases very fast and the universe evolves under a bouncing from a negative
    valued regime to the positive one. Eventually, after reaching to a maximum value at $\eta\approx 1\times10^5$, the value of $\mH$
    decreases in the expanding universe. Due to this
    form of evolution of $\mH$, we give a discussion about the strength of each term of
    $\rho_{\Lambda}$ in any epoch of the universe. As we found from Eq. \eqref{parameters},
    the parameter $n_4$ is about 38 orders of magnitude greater than $n_0$ and about 13 orders of
    magnitude greater than $n_2$. Also $n_2$ is about 25 orders of magnitude greater than $n_0$.
    Therefor at bounce point where $H=0$, the only non vanishing term is $n_0$. After that, up to $\eta\approx
    1\times10^5$, the term $n_4 H^4$ would be a dominant term, but it will
    be diluted as the universe goes on ($\mH$ decreased very fast) so that at a far future of bouncing,
    again the term $n_0$ will be dominated.

     \begin{figure}[!]
     \begin{center}
     \includegraphics[scale=0.42]{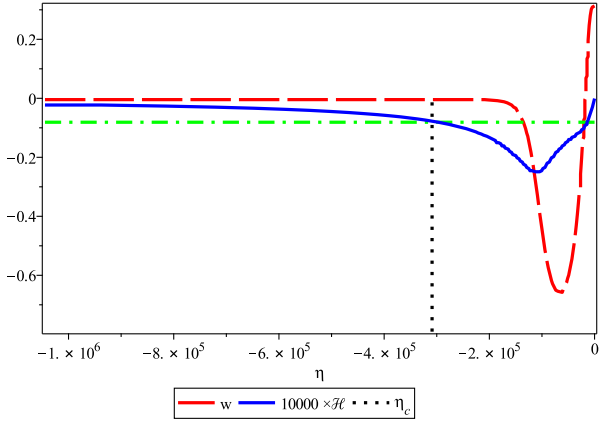}
     \caption{\footnotesize The evolution of the conformal Hubble
      parameter $\mH$ (Solid blue line) and equation of state parameter
      $w$ (dash red line), versus conformal time $\eta$ for the case contained $H^4$. Green dot-dash line is the crossing Hubble parameter
       $\mH_{cr}$ which is occurred by crossing the conformal time, $\eta_{c}$ with curve $\mH$.}
       \label{wbackground}
     \end{center}
     \end{figure}

      The evolution of the EoS parameter has been shown in fig. \ref{wbackground}
      (red dash line). As one can see, in the $\Lambda$CDM epoch, $w$ is very close to a constant
      small negative value $w \sim-0.003$. This is very good condition for getting
      a red tilt in the spectral index as indicated in \eqref{spectralxi}.
      In continuing along the conformal time about $\eta_d\approx-3.4 \times 10^5$, $w$
      decreases and after this point the time derivative of $w$ becomes negative. It is reasonable to
      expect that  after this point, dark energy has been dominated again gradually. This
      decreasing behavior is continuing until $w$ reaches to a minimum value ($w=-0.65$)
      at $ \eta \approx -6 \times 10^4$. After this point, the radiation component will be dominated.
      It should be noted that in contracting phase ($\mH<0$), Eqs. \eqref{wdot} and \eqref{running} yields
\bq
 \alpha_s\sim -\f{24 w'}{\mH},
 \eq
       which will give a negative value of spectral index for all negative values of
       $w'$ and $\mH$ at crossing time.

\begin{figure}[!]
     \begin{center}
     \includegraphics[scale=0.52]{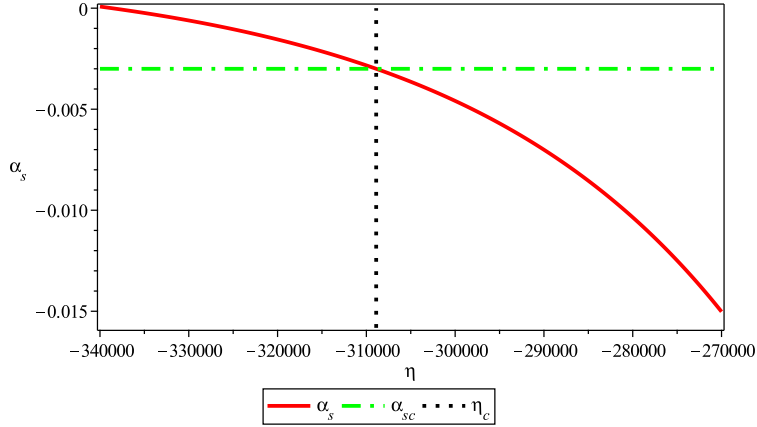}
     \caption{\footnotesize The behavior of $\alpha_s$ around the crossing time (solid line) which it is equal to $\alpha_{sc}$ at $\eta_{c}$}
     \label{alphaplot}
     \end{center}
     \end{figure}

      The evolution of the conformal Hubble parameter $\mH$ has been also shown
       (blue solid line) in fig. \ref{wbackground}. In this figure, the green horizontal dash-dot
      line shows the sound-Hubble horizon in Fourier mode $k=9.7 \times 10^{-5}$.
      Same as \cite{Cai:2014jla}, we chose the speed of sound, $c_s=0.08$ and consequently $|\mH_{cr}|=c_s k= 7.7 \times 10^{-6} $ is
      the  amount of radius of sound-Hubble horizon. Crossing time is indicated
      by vertical dot line. As it shows, this line cross the curve $\mH$ at $\eta_{c}\approx-3.1 \times 10^5
      $, which it occurs after $\eta_d$. This means that the derivative of $w$ at the crossing time
      gets a negative value. In this time, $w=w_{cr}\approx-0.0029$, $n_s=12 w_{cr}+1 \approx
      0.96$ and  after some numerical calculation we obtain $\alpha_{sc} \approx -0.003 $, which has a very good
      consistency with constrained results  ($\alpha_{sc}=-0.003 \pm 0.007 $ by $68 \%  $ CL,
       Planck+TT+LowP+Lensing \cite{Ade:2015lrj}). In fig. \ref{alphaplot}, using Eq. \eqref{alphass}, the behavior of the $\alpha_s $
       has been clearly shown around the crossing time (solid red line). As it is shown in this figure, at the horizon-crossing
       time $\eta_{c} $, the running $ \alpha_s$\ gets a negative small value ($\alpha_s \approx-0.003$,
       see the horizontal green dash-dot line in Fig.
       \ref{alphaplot}).

\subsubsection{perturbation in numerical calculation }

The evolution of cosmological perturbation is governed by Eqs.
\eqref{pertur} and \eqref{perturtensor}. We are  using a set of
parameters of \eqref{parameters}, which have been estimated in last
subsection and setting Fourier mode $k=9.6 \times 10^{-5}$. Also
during the matter dominated contracting phase, we impose the initial
conditions of the cosmological perturbation to be vacuum
fluctuation. Furthermore, we have two matter components in our model
(radiation and CDM). Since the speed of sound depends on the
background evolution, in numerical computation, approximately, we
divide the speed of the sound in two parts
 \begin{equation}
 c_s = \left\{
 \begin{array}{lr}
  \epsilon &  w\simeq-0.003\\
  \sqrt{w} &\,\,\,\, \text{radiation-dominated epoch}
 \end{array}
 \right
.
 \label{cccs}
 \end{equation}
  which $\epsilon$ is equal to $0.08$ as mentioned previously.

\begin{figure}[!h]
      \begin{center}
      \includegraphics[scale=0.6]{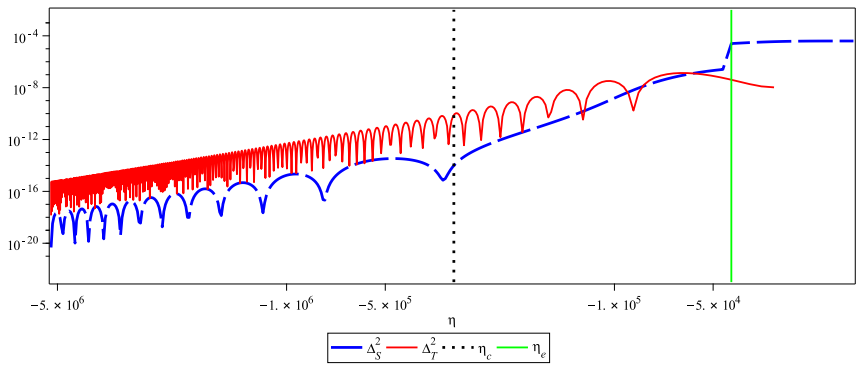}
      \caption{\footnotesize The evolution of scalar and tensor perturbation versus $\eta$ with $\epsilon=0.08$
      for the case contained term $H^4$. Dash blue line indicates
the scalar perturbation evolution and solid red line indicates the
tensor perturbation evolution. Dot black line indicates the crossing
time and vertical solid green line is referred to the equality
time.}
       \label{conformalHper}
        \end{center}
        \end{figure}

 The evolution of the scalar cosmological perturbation (blue dash line)
 has been depicted in fig. \ref{conformalHper}. As it can be seen, curvature
 perturbation oscillates from the sub-Hubble
  scale to super-Hubble region at $\eta_{c}\approx-3.1 \times 10^5$, in which the oscillation finished.
  In fact $k$-mode curvature perturbation exits from the sound Hubble horizon at the crossing point $\eta_{c}$.
  As it shows, after equality time (vertical green line) when $\rho_r= \rho_m+\rho_{\Lambda}$, we can
  consider the radiation begins to dominate or in analytically point of view, $ {z''}/{z} \rightarrow 0$, and consequently the amplitude of the
  scalar perturbation approximately becomes constant.

 The red solid line shows the oscillation of the tensor cosmological perturbation or
 gravitational wave. The sound speed of the tensor perturbation is $c_s^T=1$. So
 clearly, tensor perturbation continue to oscillates even after vertical black dot
 line and eventually will damp to a constant value after equality time.

 Besides the nearly scale invariant power spectrum with a negative running, a small tensor to scalar
  ratio is predicted by a cosmological bounce scenario.  The
  tensor-to-scalar ratio $r$ is constrained by the observational bound ($ r < 0.12$) \cite{Ade:2015tva}.
  There are some known mechanisms for predicting a small tensor to scalar ratio
  \cite{Cai:2013kja,Cai:2011zx,Fertig:2016czu,Wilson-Ewing:2015sfx}.

\begin{figure}[h]

      \begin{center}
      \includegraphics[scale=0.4]{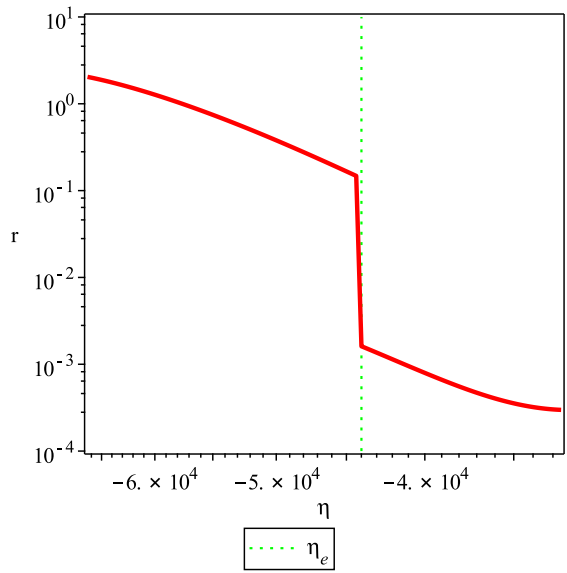}
      \caption{\footnotesize The evolution of tensor to scalar  ratio ($r$) versus $\eta$ for the case contained term $H^4$. The ratio at
      $\eta_e$ suddenly decreased to $r\leq0.001$. }
       \label{r}
        \end{center}
        \end{figure}

   Actually, in our model, the speed of sound in matter dominated epoch is less than unity,
   so it affects on the amplitude of
   the scalar perturbations and consequently the amplitude of the scalar
   perturbation can be increased against the amplitude of the tensor perturbation. Also as it is shown in fig.
   \ref{conformalHper}, the discontinuity of $c_s$ about the equality time, suddenly increased the amplitude of the scalar perturbation.
   The evolution of tensor to scalar ratio $r$ is shown in fig. \ref{r}. As one can see, before
   the equality time (dot green line), the amount of tensor to scalar ratio $r$ slowly decreases from  high to low.

     In order to compare all studied cases of RVM-bounce scenario, we summarize all of them in
      table \ref{chartt}.
      As we found, for the cases 1-4, some physical requirements such as $\rho_{\Lambda}>0$ (or $n_0>0$)
      at any time and $w<0$ at the crossing time, required that $\alpha_s>0$. In the final column of the table \ref{chartt}, the value of $\alpha_{sc}$ is
      calculated for all cases at crossing time, where $w=w_{cr}=-0.003$, and $H=H_{cr}=-8.8 \times 10^{-8}$ for $n_0>0$.
       At last, except the case contained the term $H^4$, other cases cannot satisfy the weak energy condition and observational
       evidence simultaneously. Therefore we excluded them in further numerical calculation analysis (such as tensor to scalar ratio $r$).

\begin{table*}[!h]
\centering

\resizebox{\columnwidth}{!}{
\begin{tabular}{|c|c|c|c|c|c|}

\hline

  &$\rho_{\Lambda}$ & $ w =(n_s-1)/12$ &  $\dot{w}$ &$\alpha_s$  & $\alpha_{sc}$\\
\hline\hline
\Bigg| 1 &$n_0$ & $\dfrac{-n_0}{3H^2}$& $3 H w(1+w)$ & $-72 w$& 0.22\\
\hline
 \Bigg| 2 &$n_0+n_2 H^2$ & $ -\dfrac{1}{3}(\dfrac{n_0}{H^2}+n_2)$ & $\dfrac{3 n_0H}
 {n_0+n_2 H^2}w(1+w)$ &$\dfrac{n_0}{n0+n_2H^2}(-72 w) $& 0.22 \\
\hline \Bigg| 3 &$n_0+n_1 H$ &$-\dfrac{1}{3}(\dfrac{n_0}{
H^2}+\dfrac{n_1}{ H})$&
$\dfrac{ 3 H (n_1 H +2 n_0)}{2 n_0 +n_1 H}w(1+w)$& $\dfrac{n_1 H+2 n_0}{n_1 H+n_0}(-72 w)$& 0.44\\

\hline

\Bigg| 4 &$n_0+\beta \dot{H}$ &$ \dfrac{2 n_0/H^2-3
\beta}{-6+3\beta}$&
$\dfrac{3 H n_0 w (1+w)}{n_0-3\beta H^2/2} $& $\dfrac{n_0}{n_0-3 \beta H^2 /2} (-72 w)$& 0.001 \\

\hline

\bigg| 5 & $n_2 H^2 +\beta \dot{H}$& $ \dfrac{-2 n_2+3 \beta}{ 6-3
\beta}$
& 0& 0& 0\\
\hline

\Bigg| 6 &$n_0+n_2 H^2 +n_4 H^4$ &$
-\dfrac{1}{3}(\dfrac{n_0}{H^2}+n_2+n_4 H^2)$&
$ \dfrac{3 H( -n_4 H^4 +n_0)w(1+w)}{n_0+n_2 H^2 +n_4 H^4} $&$ \dfrac{(-n_4 H^4+n_0)(-72 w)}{n_0+n_2 H^2 +n_4 H^4} $ & -0.003\\

\hline

\end{tabular}

} \caption{\footnotesize Summary of all cases in RVM-bounce scenario
. Note that in all cases, $w$ has a small negative value (see Eq.
(\ref{ww})). Also the running $\alpha_{sc}$ is the value of
$\alpha_s$ at crossing, $H=H_{cr}=- 8.8\times10^{-8}$ (in
contracting phase) and $w=w_{cr}=-0.003$ provided that
$\rho_{\Lambda}>0$ even at the bounce point
($\mH=0$).}\label{chartt}
\end{table*}

 \section{conclusion}

In this work we introduced a deformed matter bounce scenario with
the running vacuum model (RVM). This model could be considered as a
viable alternative to the inflationary paradigm both in
observational and theoretical aspects. Based on RVM-DE, the standard
cosmological constant not more constant, but may consider as series
of powers of $H^2$ and $\dot H$. By introducing some cases of RVM,
we calculated the spectral index $n_s$ and its running $\alpha_s$ in
order to compare with observational data. In fact in the contracting
phase, before a bouncing, when the EoS parameter is slightly
negative, Fourier modes of perturbations exit from the sound
horizon. Thus power spectrum of cosmological perturbation for long
wavelength modes is not exactly scale invariant and consequently it
gets a slightly red tilt.

The process of creating of red tilt is obviously indicated in the
analytical treatment for the case  $\Lambda=n_2 H^2 + \alpha
\dot{H}$. In this case the running of spectral
 index become vanishing, $\alpha_s=0$, which is inconsistent with inflationary paradigm. Some models with
 expansion up to $H^2$ got positive running and for a model $\Lambda(H)=n_0+n_2 H^2+n_4 H^4$,
 by estimating a set of parameters, we obtained the spectral
index $n_s\approx 0.96$, running of spectral index $\alpha_s<0$ and
tensor to scalar ratio $r<0.12$. We found that this model had the
best consistency with the cosmological observations and reveals a
degeneracy between deformed matter bounce scenario with RVM-DE and
inflation. As a work in the future, the observational constraint of
this model and comparison with cosmological observations are
suggested. 
\bibliography{ref}
\end{document}